\documentclass[aps,twocolumn,superscriptaddress,preprintnumbers]{revtex4}
\usepackage{graphicx}

\begin{document}

\preprint{IRB-TH-1/02}

\title{IR finite
       one$-$loop box scalar integral
       with massless internal lines}


\author{G. Duplan\v ci\' c}
\author{B. Ni\v zi\' c}
\affiliation{Theoretical Physics Division, Rudjer Bo\v{s}kovi\'{c} Institute, 
        P.O. Box 180, HR-10002 Zagreb, Croatia}


\begin{abstract}
The IR finite one-loop box scalar integral 
with massless internal lines has been recalculated. 
The result is very compact, simple and valid for
arbitrary values of the relevant kinematic variables. It is given in terms 
of only two dilogarithms and a few logarithms, all of very simple arguments.
\end{abstract}


\maketitle

\newcommand{\ri}{{\rm i}}
\newcommand{\e}{{\rm i}\epsilon}
\newcommand{\nn}{\nonumber}

\section{Introduction}
\label{intro}
Exclusive hadronic processes at large momentum transfer in which the total of
particles (partons) in the initial and final states is $N\geq 5$ are becoming
increasingly important in testing various aspects of QCD.
The tree-level results for the cross section of these processes, being
proportional to a high power of the QCD coupling constant $\alpha_S^N$, are very
sensitive to the variation of the renormalization scale and scheme being used.
Consequently, to stabilize the leading-order (tree-level) predictions and to
achieve a complete confrontation between theoretical predictions and
experimental data, at least one-loop (NLO) corrections are necessary.

The main theoretical difficulty in obtaining the NLO predictions consists in the
treatment of the occurring $N$-point one-loop
scalar and tensor integrals with massless
internal lines.
Through the scalarization and reduction procedures \cite{dixon,tarasov,binoth},
the computation of such integrals reduces to the calculation of a set comprised
of six box ($N=4$) diagrams, one of which is finite while the other five are IR
divergent.
Being of fundamental importance for one-loop calculations in pQCD with
massless quarks, it is very important for practical purposes that the results
for these integrals are obtained in forms as compact as possible (expressed in
terms of as fewer as possible dilogarithmic functions) with as simple as
possible arguments.

The IR divergent basic box integrals have been considered in 
Refs. \cite{fabricius,papa,bern}. A more general and detailed analysis has been
given in Ref. \cite{dupli}.
As for the IR finite box integral, it has been calculated in
Ref. \cite{hooft}, and later obtained in a more compact form in 
Ref. \cite{denner}. From the practical point of view, in addition to not being
most compact, a disadvantage
of this result
is a very complicated structure of the arguments of the occurring
dilogarithmic and logarithmic functions.
An alternative expression for the IR finite box integral has been obtained in 
Ref. \cite{ussyukina}, where it has been related to the IR
finite triangle scalar integral.
The result of Ref. \cite{ussyukina} expressed in terms of only two
dilogarithms, however, is not valid for arbitrary values
of the relevant kinematic variables.

Making use of the method of Ref. \cite{ussyukina}, in this work we carefully
recalculate the IR finite box scalar integral.
Our result is as simple and compact as that of 
Ref. \cite{ussyukina} and, as the result of Refs. \cite{hooft,denner}
 is quite general, i.e. applicable to all kinematic
regions.
In addition, to be more compact and general, 
the advantage of our result over the
results
previously obtained is also the fact that one can very easily separate
the real and imaginary parts of logarithms and dilogarithms, making it more
appropriate for numerical calculations.

The paper is organized as follows. Sect. 2 is devoted to introducing the
notation and to some preliminary considerations. In Sect. 3, using the Feynman 
parameter method and the Mellin$-$Barnes integral representations, we evaluate
the IR finite box scalar
integral by relating it to the IR finite
triangle scalar integral. In Sect. 4 we give some concluding remarks. 
The details of the evaluation of the IR finite triangle scalar 
integral are given in the
Appendix.

\section{Preliminaries}
\label{sec:pre}
\begin{figure}
 \resizebox{0.5\textwidth}{!}{%
  \includegraphics{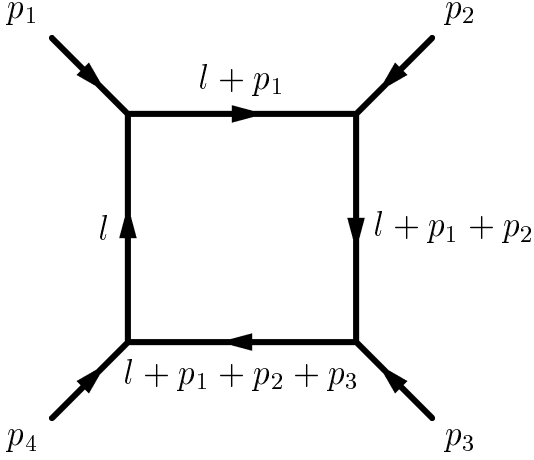}
}
 \caption{One-loop box diagram.}
 \label{f:box}
\end{figure}

The scalar one$-$loop box integral with massless internal lines in
$D$$-$di\-men\-si\-o\-nal
space$-$time is given by
\begin{equation}
I_{4}(p_1,p_2,p_3,p_4)=({\mu}^2)^{2-D/2}
\int \frac{{\rm d}^{D}l}{(2\pi)^{D}}\frac{1}
{A_1 A_2 A_3 A_4}~,\label{eq:f1}
\end{equation}
where $p_i$ ($i$=1,2,3,4) are the external momenta, $l$ is the loop
momentum and $\mu$ is the usual dimensional regularization scale.
As indicated in Fig. \ref{f:box}, all external momenta are taken to
 be incoming,
so that the massless propagators are
\begin{eqnarray}
A_{1} &=& l^{2}+{\rm i}\epsilon \,,\nonumber \\
A_{2} &=& (l+p_{1})^{2}+{\rm i}\epsilon \,,\nonumber \\
A_{3} &=& (l+p_{1}+p_{2})^{2}+{\rm i}\epsilon \,,\nonumber \\
A_{4} &=& (l+p_{1}+p_{2}+p_{3})^{2}+{\rm i}\epsilon \,,\label{eq:f2} 
\end{eqnarray}
where the quantity $\e$ ($\epsilon >0$), represents an infinitesimal imaginary
part specifying on which side of the cut a multivalued function has to be
evaluated. We take the cut along the negative real axis.

Combining the denominators using the Feynman 
pa\-ra\-me\-tri\-za\-ti\-on formula,
performing the $D$$-$dimensional loop momentum integration,
introducing the external "masses"
$p_i^2=m_i^2$ $(i=1,2,3,4)$
and the Mandelstam variables
$s=(p_1+p_2)^2$ and $t=(p_2+p_3)^2$, the integral
(\ref{eq:f1}) becomes
\begin{eqnarray}
\lefteqn{I_4(s,t;m_1^2,m_2^2,m_3^2,m_4^2)= 
\frac{{\rm i}}{(4\pi)^2}
\frac{\Gamma (4-D/2)}
{(4\pi {\mu}^2)^{D/2-2}}}\nonumber \\
& &\times \int_0^1{\rm d}x_1 {\rm d}x_2 {\rm d}x_3 {\rm d}x_4 
\;\delta(x_1+x_2+x_3+x_4-1)
\nonumber \\
& &\times \left(-x_1x_3\;s-x_2x_4\;t-x_1x_2\;m^2_1\right . \nonumber \\
& &{}\left .-x_2x_3\;m^2_2
-x_3x_4\;m^2_3-x_1x_4\;m^2_4-{\rm i}\epsilon\right)^{D/2-4}. \label{eq:f3}
\end{eqnarray}
Depending on the number of external massless lines, there are
six special cases of the above integral.
Adopting the notation of Ref. \cite{bern}, we denote them by
\begin{eqnarray}
I_4^{0m} &\equiv & I_4(s,t;0,0,0,0,), \label{eq:f4a} \\
I_4^{1m} &\equiv & I_4(s,t;0,0,0,m_4^2), \label{eq:f5a} \\
I_4^{2me} &\equiv & I_4(s,t;0,m_2^2,0,m_4^2), \label{eq:f6a} \\
I_4^{2mh} &\equiv & I_4(s,t;0,0,m_3^2,m_4^2), \label{eq:f7a} \\
I_4^{3m} &\equiv & I_4(s,t;0,m_2^2,m_3^2,m_4^2), \label{eq:f8a} \\
I_4^{4m} &\equiv & I_4(s,t;m_1^2,m_2^2,m_3^2,m_4^2). \label{eq:f9a} 
\end{eqnarray}
The above integrals constitute the fundamental
set of integrals, in the sense that an arbitrary one$-$loop $N(\ge 5)-$po\-int
integral with massless internal lines can be uniquely expressed
as
a linear combination of these integrals with the coefficients being
rational functions of the relevant kinematic variables.
The integrals (\ref{eq:f4a}$-$\ref{eq:f8a}) are IR divergent and need to be
evaluated in $D=4+2 \varepsilon_{IR}$ ($\varepsilon_{IR}>0$) dimensions
\cite{dupli}. On the other hand, the integral (\ref{eq:f9a}) 
is IR finite and can
be calculated in $D=4$ dimensions.

Setting $D=4$ in (\ref{eq:f3}), we obtain
\begin{eqnarray}
& &I_4(s,t,m_1^2,m_2^2,m_3^2,m_4^2)=\frac{\ri}{(4\pi)^2}
\int_0^1\!\! {\rm d}x_1\, {\rm d}x_2\, {\rm d}x_3\, {\rm d}x_4\nn \\
& &\hspace{-2.5mm}{}\times \delta(x_1+x_2+x_3+x_4-1) \left(x_1x_3\;s+x_2x_4\;t
 \right.\nn \\
& &\hspace{-2.5mm}\left.{}+x_1x_2\;m^2_1+x_2x_3\;m^2_2+x_3x_4\;m^2_3
+x_1x_4\;m^2_4+\e
\right)^{-2}.\nn \\
& &\label{eq:f6}
\end{eqnarray}
This integral is invariant under some permutations of external "masses" and
Mandelstam variables.
Namely, introducing the following set of ordered pairs
\begin{equation}
\{\,(s,t), (m_1^2,m_3^2), (m_2^2,m_4^2 )\,\},\label{eq:f7}
\end{equation}
one can easily see that the integral (\ref{eq:f6}) is invariant under the
permutations of ordered pairs given in (\ref{eq:f7}),
 as well as under the simultaneous
exchange of places of elements in any two pairs.
Furthermore, the integral is invariant under the simultaneous change 
of the signs of the
kinematic variables and the causal $\e$. 
Consequently, it is sufficient to analyze only the cases
of the integral (\ref{eq:f6})  
in which the number of positive kinematic variables is larger than the number
of negative ones.

\section{Calculation and results}
\label{sec:four}
Eliminating the $\delta-$function in (\ref{eq:f6})
by performing the $x_4$ integration, we get
\begin{eqnarray}
\lefteqn{\hspace{-3mm}I_4(s,t;m_1^2,m_2^2,m_3^2,m_4^2)=\frac{\ri}{(4\pi)^2}
\!\int_0^1\!\!{\rm d}x_1 \!\int_0^{1-x_1}\!\!\hspace{-5mm}{\rm d}x_2
  \!\int_0^{1-x_1-x_2}\!\!\hspace{-5mm}{\rm d}x_3}\nn \\
& & \times \big[ (1-x_1-x_2-x_3)(\,x_1 \,m_4^2+x_2 \,t+x_3 \,m_3^2\,)
\hspace{5mm}\nn \\
& &{}+x_1 x_2\,m_1^2+x_1 x_3\,s+x_2 x_3\,m_2^2
+\e  \, \big]^{-2}.\label{eq:f38}
\end{eqnarray}
Next, if we first make the substitution $x_3\to (1-x_1-x_2) x_3$ and then  
$x_2\to (1-x_1) x_2$, we find that
\begin{eqnarray}
\lefteqn{I_4(s,t;m_1^2,m_2^2,m_3^2,m_4^2)=\frac{
\ri }{(4\pi)^2}
  \int_0^1\!\!\!{\rm d}x_1 \int_0^{1}\!\!\!{\rm d}x_2
  \int_0^{1}\!\!\!{\rm d}x_3}\nn \\
& &{}\times(1-x_2)\bigg\{(1-x_1)(1-x_2)\left[x_2 x_3\,m_2^2
+x_2 (1-x_3)\,t\right.\nn \\
& &{}+\left.
(1-x_2)x_3 (1-x_3)\,m_3^2 \right]+x_1 \left[x_2\,m_1^2+(1-x_2) x_3\,s\right.\nn \\
& &{}+\left.(1-x_2)(1-x_3)\,m_4^2
 \right]+\e \bigg\}^{-2}.\label{eq:f39}
\end{eqnarray}
Performing the $x_1$ integration and a few simple
rearrangements, the integral takes
the form
\begin{eqnarray}
\lefteqn{I_4(s,t;m_1^2,m_2^2,m_3^2,m_4^2)=
\frac{
\ri }{(4\pi)^2}
\int_0^{1}\!\!\!{\rm d}x_2\int_0^{1}\!\!\!{\rm d}x_3
}\nn \\
& &\hspace{-5mm}\times \big[x_2\,m_1^2+(1-x_2) x_3\,s+(1-x_2)(1-x_3)\,m_4^2
+\e  \big]^{-1}\nn \\
& &\hspace{-5mm}\times \big[x_2 x_3\,m_2^2
+x_2 (1-x_3)\,t+(1-x_2)x_3 (1-x_3)\,m_3^2 +\e \big]^{-1}\hspace{-2mm}.\nn
\\ & & \label{eq:f40}
\end{eqnarray}
Next, doing the $x_2$ integration and expressing the remaining integral in
terms of logarithms and dilogarithms, one arrives at the result obtained in
Ref. \cite{denner}, which is valid for all values of the kinematic variables.
The result of Ref. \cite{denner} is given in terms
of logarithmic and dilogarithmic functions with
the argument of the form
\[
1+\frac{f_1-\e}{f_2-\e}\left( f_3\pm\sqrt{f_4+{\rm i}f_5\epsilon}\right),
\]
where $f_i$ $(i=1,2,3,4,5)$ are real functions of the kinematic variables.
As seen from the above, the structure of the argument is very complicated
since $\e$ appears both outside and inside the square root. There are two
consequences of this: first, it is very dificult to determine the sign of the
infinitesimal imaginary part of the argument; second, it is extremely
complicated to use
the dilogarithmic identities with the aim of simplifying the final result. 
It should also be pointed out that since the final result contains products of
logarithms, the signs of the imaginary parts of arguments are important
not only for determining the imaginary part of the result, but also
for
the real part of the result.
From the practical point of view, this represents a disadvantage of the result for
the IR finite massless box scalar integral derived in Ref. \cite{denner}.

To avoid the above-mentioned problems, we take a different approach and proceed
to evaluate the box integral by expressing it using the Mellin-Barnes integral
representation \cite{ussyukina}.
To this end, we first need to find the Mellin-Barnes representation of the
following expression:
\begin{equation}
J=(\,y_1 y_2\,a+y_1 (1-y_2)\,b+(1-y_1) y_2 (1-y_2)\,c+\e\,)^{-1}, \label{eq:f41}
\end{equation}
where $y_1,\,y_2\in [0,1]$, and $a$, $b$ and $c$ are arbitrary real
variables different from zero. 
In what follows we also need the Mellin-Barnes
 representation of the expression $(1-z)^{-k}$
\begin{eqnarray}
(1-z)^{-k} &=& \frac{1}{2\pi \ri \Gamma(k)}
\int_{\gamma-\ri \infty}^{\gamma+\ri \infty} 
\!\!\!{\rm d}s\, \Gamma(s) \Gamma(k-s)
(-z)^{-s} \nn \\
& & 0<\gamma<{\rm Re}\,k, \qquad |{\rm arg}(-z)|<\pi. \label{eq:f42}
\end{eqnarray}
Rewriting the expression (\ref{eq:f41}) as 
\begin{eqnarray}
J &=& \frac{1}{y_1 [y_2\,a+(1-y_2)\,b+\e]}\nn \\
& &\times \left(\,1+\frac{1-y_1}{y_1}
\frac{y_2 (1-y_2)\,c+\e}{y_2\,a+ (1-y_2)\,b+\e}\right)^{-1}, \label{eq:f43}
\end{eqnarray}
and utilizing the relation (\ref{eq:f42}), $J$ can be written in the form
\begin{eqnarray}
J &=& \frac{1}{2\pi \ri}
\int_{\gamma-\ri \infty}^{\gamma+\ri \infty} \hspace{-3mm}{\rm d}s 
\,\,\frac{\Gamma(s) 
\Gamma(1-s)\,y_1^{s-1}}{[(1-y_1)(y_2 (1-y_2)\,c+\e)]^{s}}\nn
\\
& &\times [\,y_2\,a+ (1-y_2)\,b+\e\,]^{s-1},
\hspace{4mm}0 < \gamma < 1.   \label{eq:f44}
\end{eqnarray}
Now, applying the relation (\ref{eq:f42})
to the factor $(y_2\,a+ (1-y_2)\,b+\e)^{s-1}$,  appearing on the right-hand
side of
(\ref{eq:f44}), one finds that
\begin{eqnarray}
J &=& \frac{1}{(2\pi \ri)^2} \frac{1}{a+\e}
\int_{\gamma-\ri \infty}^{\gamma+\ri \infty} \hspace{-5mm}{\rm d}s
\int_{\gamma'-\ri \infty}^{\gamma'+\ri \infty} \hspace{-5mm}{\rm d}s' 
\, \Gamma(s) \Gamma(s') \Gamma(1-s-s')\nn \\
& &\times \, \frac{y_1^{s-1} y_2^{s'-1}}
{(1-y_1)^{s}(1-y_2)^{s+s'}} \left(\frac{c+\e}{a+\e}\right)^{\!\!-s}
 \left(\frac{b+\e}{a+\e}\right)^{\!\!-s'}\hspace{-3mm},
\nn \\
& &\hspace{16mm} 0 < \gamma ,\gamma'< 1,  \qquad \gamma +\gamma'< 1,
\label{eq:f45}
\end{eqnarray}
which represents the desired Mellin-Barnes representation 
for $J$ given in (\ref{eq:f41}).
If use is made of the relation
(\ref{eq:f45}), the integral (\ref{eq:f40}) becomes
\begin{eqnarray}
\lefteqn{I_4(s,t;m_1^2,m_2^2,m_3^2,m_4^2)=
\frac{
\ri }{(4\pi)^2}
\frac{1}{(2\pi \ri)^2} \frac{1}{m_2^2+\e}} \nn \\
& &\times \int_{\gamma-\ri \infty}^{\gamma+\ri \infty} \hspace{-3mm}{\rm d}s
\int_{\gamma'-\ri \infty}^{\gamma'+\ri \infty} \hspace{-3mm}{\rm d}s' 
\, \Gamma(s) \Gamma(s')\,\Gamma(1-s-s')\nn \\
& &\times \left(\frac{m_3^2+\e}{m_2^2+\e}\right)^{\!\!-s}
 \left(\frac{t+\e}{m_2^2+\e}\right)^{\!\!-s'}
 \!\!\int_0^{1}\!\!\!{\rm d}x_2\int_0^{1}\!\!\!{\rm d}x_3\nn \\
& &\times 
\frac{x_2^{s-1} (1-x_2)^{-s} x_3^{s'-1}
 (1-x_3)^{-s-s'}}{x_2\,m_1^2+(1-x_2) x_3\,s+(1-x_2)(1-x_3)\,m_4^2
+\e}.\nn \\& & \label{eq:f46}
\end{eqnarray}
After performing the $x_2$ and $x_3$ integrations with the help of the formula
\begin{eqnarray}
\lefteqn{\hspace{-10mm}\int_0^1{\rm d}x\, x^{b-1} (1-x)^{a-b-1} (1-x\,z)^{-a}=
\frac{\Gamma(b)\Gamma(a-b)}{\Gamma(a)\,(1-z)^b},}\nn \\
& &  {\rm Re}\,a > {\rm Re}\,b > 0,\,\,\,|{\rm arg}(1-z)|<\pi,
\label{eq:f47}
\end{eqnarray}
one arrives at
\begin{eqnarray}
\lefteqn{\hspace{-2mm}I_4(s,t;m_1^2,m_2^2,m_3^2,m_4^2)=
\frac{
\ri }{(4\pi)^2}
\frac{1}{(2\pi \ri)^2} \frac{1}{(m_2^2+\e)(m_4^2+\e)}} 
\nn \\
& &\times\int_{\gamma-\ri \infty}^{\gamma+\ri \infty} \hspace{-3mm}{\rm d}s
\int_{\gamma'-\ri \infty}^{\gamma'+\ri \infty}\hspace{-3mm} {\rm d}s'
 \left[\Gamma(s) \Gamma(s') \Gamma(1-s-s')\right]^2\,\nn \\
& &\times \frac{(m_1^2+\e)^{-s}}{(m_2^2+\e)^{-s}}
  \frac{(m_3^2+\e)^{-s}}{(m_4^2+\e)^{-s}}\frac{(s+\e)^{-s'}}{(m_2^2+\e)^{-s'}}
\frac{(t+\e)^{-s'}}{(m_4^2+\e)^{-s'}},\nn \\
& & \qquad \qquad 
0 < \gamma ,\gamma'< 1,  \qquad \gamma +\gamma'< 1,\label{eq:f50}
\end{eqnarray}
which is the sought Mellin-Barnes representation of the IR finite box scalar
integral.

\begin{figure}
 \resizebox{0.5\textwidth}{!}{%
  \includegraphics{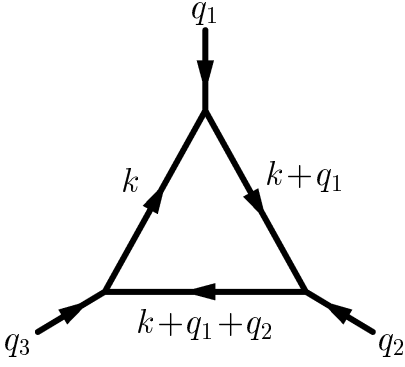}
}
 \caption{One-loop triangle diagram.}
 \label{f:tri}
\end{figure}

Let us now make a digression and, for a moment, consider the IR finite scalar
one-loop triangle integral with massless internal lines
in 4-dimensional space-time. 
It corresponds to
the Feynman diagram of Fig. \ref{f:tri}, and is given by
\begin{eqnarray}
\lefteqn{I_{3}(q_1,q_2,q_3)=}\label{eq:A1} \\
& &\int \frac{{\rm d}^{4}k}{(2\pi)^{4}}\frac{1}
{[k^{2}+\e]\,[(k+q_{1})^{2}+\e]\,[(k+q_{1}+q_{2})^{2}+\e]}~,\nn
\end{eqnarray}
where $q_i$, $i$=1,2,3 are the incoming 
external momenta.
Upon combining the denominators with the help of the Feynman parametrization 
formula, integrating out the loop momentum and introducing the 
external "masses"
$q_i^2=\mu_i^2~~(i=1,2,3)$, 
the integral (\ref{eq:A1}) takes the form
\begin{eqnarray}
I_3(\mu_1^2,\mu_2^2,\mu_3^2)&=&\frac{
\ri }{(4\pi)^2}\!\!
\int_0^1\!\!\! 
{\rm d}x_1\,  {\rm d}x_2\, {\rm d}x_3\, \delta(x_1+x_2+x_3-1)\nn \\
& &{}\hspace{-1.5cm}\times(x_1 x_2\;\mu_1^2+x_2 x_3\;\mu_2^2 +x_1 x_3\;\mu_3^2
+\e)^{-1}.\label{eq:A2}
\end{eqnarray}
In this form it is evident that the triangle integral is invariant
under arbitrary permutations of external masses $\mu_i^2$.

After eliminating the $\delta$-function by performing the $x_3$ integration,
introducing the new variable $x_2\to (1-x_1) x_2$ and taking into account Eq.
(\ref{eq:f45}), one can write
\begin{eqnarray}
\lefteqn{I_3(\mu_1^2,\mu_2^2,\mu_3^2)=\frac{
\ri }{(4\pi)^2}
\frac{1}{(2\pi \ri)^2} \frac{1}{\mu_1^2+\e}
\int_{\gamma-\ri \infty}^{\gamma+\ri \infty} \hspace{-3mm}{\rm d}s
\int_{\gamma'-\ri \infty}^{\gamma'+\ri \infty} \hspace{-3mm}{\rm d}s'
} 
 \nn \\
& &\times \Gamma(s) \Gamma(s') \Gamma(1-s-s')
\left(\frac{\mu_2^2+\e}{\mu_1^2+\e}\right)^{\!-s}
 \left(\frac{\mu_3^2+\e}{\mu_1^2+\e}\right)^{\!-s'}\nn \\
& &\times \int_0^1\! 
 {\rm d}x_1 x_1^{s-1} (1-x_1)^{-s} \int_0^{1}\!{\rm d}x_2
 x_2^{s'-1} (1-x_2)^{-s-s'}\!\!\!.\nn \\ & &\label{eq:f51}
\end{eqnarray}
Next, performing the $x_1$ and $x_2$ integrations with the help of the formula
(\ref{eq:f47}) (with $z=0$), one obtains
\begin{eqnarray}
\lefteqn{I_3(\mu_1^2,\mu_2^2,\mu_3^2)=\frac{
\ri }{(4\pi)^2}
\frac{1}{(2\pi \ri)^2} \frac{1}{\mu_1^2+\e} 
\int_{\gamma-\ri \infty}^{\gamma+\ri \infty} \hspace{-3mm}{\rm d}s
\int_{\gamma'-\ri \infty}^{\gamma'+\ri \infty} \hspace{-3mm}{\rm d}s'} \nn \\
& & \times
\left[ \Gamma(s) \Gamma(s') \Gamma(1-s-s')\right] ^2
\frac{(\,\mu_2^2+\e\,)^{-s}}{(\,\mu_1^2+\e\,)^{-s}}
\frac{(\,\mu_3^2+\e\,)^{-s'}}{(\,\mu_1^2+\e\,)^{-s'}},\nn \\
& & \qquad \qquad 
0 < \gamma ,\gamma'< 1,  \qquad \gamma +\gamma'< 1,
\label{eq:f52}
\end{eqnarray}
which represents the Mellin-Barnes representation of the triangle 
scalar integral.

If we now compare the Mellin-Barnes representations of the scalar box and
triangle integrals given by 
Eqs. (\ref{eq:f50}) and (\ref{eq:f52}), respectively, we arrive at the
relation
\begin{equation}
I_4(s,t;m_1^2,m_2^2,m_3^2,m_4^2)=I_3(s t,m_1^2 m_3^2,m_2^2 m_4^2),
\end{equation}
which is valid provided the following condition is satisfied:
\begin{equation}
(u+\e)^{-s''}(v+\e)^{-s''}=(uv+\e)^{-s''},\label{cond}
\end{equation}
where $(u,v)$ is any element of the set of ordered pairs in
 (\ref{eq:f7}) and $s''$ stands for $s$ or $s'$.
As is easily seen,
the above condition does not hold only if $u$ and $v$
are both negative ($u,\,v <0$). Since, as mentioned earlier, 
we are considering the
cases with more positive than negative
parameters, it is clear that 
the case when (\ref{cond}) is not valid can appear at most in one 
of the pairs.

Let us now examine in detail the product appearing on the 
left-hand side in
(\ref{cond}), corresponding to the case when
$u,\,v <0$.
One can then write
\begin{eqnarray}
\lefteqn{(u+\e)^{-s''}(v+\e)^{-s''}=|u|^{-s''}|v|^{-s''} 
\exp [-2\ri \pi\, {\rm sign}(\epsilon)\,s'']}\nn \\
&=& (uv)^{-s''}[\cos (\pi\, s'')-
\ri\,{\rm sign}(\epsilon)\sin (\pi\, s'')]^2\nn \\
&=& (uv)^{-s''}[1-2\ri\,{\rm sign}(\epsilon)\sin (\pi\, s'')
\cos (\pi\, s'')\nn \\
& &{}-2\sin ^2(\pi\, s'')]\nn \\
&=&(uv)^{-s''}\{1-2\ri\,{\rm sign}(\epsilon)\sin (\pi\, s'')
\exp [-\ri \pi\, {\rm sign}(\epsilon)\,s'']\}\nn \\
&=&(uv+\e)^{-s''}-2\ri\,{\rm sign}(\epsilon)\sin (\pi\, s'')\,(-uv+\e)^{-s''}
.\label{eq:f53a} 
\end{eqnarray}
Next, with the help of the formula
\[ \Gamma(z) \Gamma(1-z)=\pi \sin^{-1}(\pi\, z),\]
one finds that
\begin{eqnarray}
\lefteqn{(u+\e)^{-s''}(v+\e)^{-s''}=
(uv+\e)^{-s''}}\nn \\
& &{}-\frac{2\ri\pi\,{\rm sign}(\epsilon)}{\Gamma(s'') \Gamma(1-s'')}
(-uv+\e)^{-s''}.
\label{eq:f53} 
\end{eqnarray}
Owing to the symmetry of the pairs, with no loss of generality, we can now
suppose that
the "problematic pair" (i.e. the pair in which both masses are negative) 
is $(m_1^2,m_3^2)$. 
In this case, the first term on the right-hand side in 
(\ref{eq:f53}) is of the form that makes it possible to 
represent the box integral in terms of the 
triangle integral, while the second term is a correction 
originating from the fact
that the first term does not keep the information regarding the
signs of  $u$ and $v$.

Let us now evaluate the correction term. 
Upon combining Eqs. (\ref{eq:f50}) and (\ref{eq:f53}),  
the Mellin-Barnes representation of the correction term is
\begin{eqnarray}
\lefteqn{K(s, t, m_1^2, m_3^2, m_2^2, m_4^2)=\frac{
\ri }{(4\pi)^2}
\frac{1}{(2\pi \ri)^2} 
\frac{-2\ri\pi\,{\rm sign}(\epsilon)}{m_2^2m_4^2+\e}}\nn \\
& &\times\int_{\gamma-\ri \infty}^{\gamma+\ri \infty} \hspace{-3mm}{\rm d}s
\int_{\gamma'-\ri \infty}^{\gamma'+\ri \infty} \hspace{-3mm}{\rm d}s' 
\frac{\Gamma(s) \left[\Gamma(s') \Gamma(1-s-s')\right]^2}{\Gamma(1-s)}
 \nn \\
& &\times \frac{(-m_1^2m_3^2+\e)^{-s}}{(m_2^2m_4^2+\e)^{-s}}
\frac{(st+\e)^{-s'}}{(m_2^2m_4^2+\e)^{-s'}}
 \label{eq:f54} 
\end{eqnarray}
which, after making use of the formula (\ref{eq:f47}) (with $z=0$,
$a=1-s$ and $b=s'$), is found to
take the form
\begin{eqnarray}
\lefteqn{K(s, t, m_1^2, m_3^2, m_2^2, m_4^2)=\frac{
\ri }{(4\pi)^2}
\frac{1}{(2\pi \ri)^2} 
\frac{-2\ri\pi\,{\rm sign}(\epsilon)}{m_2^2m_4^2+\e}}
\nn \\
& &\times \int_0^1\hspace{-1mm}{\rm d}y
\int_{\gamma-\ri \infty}^{\gamma+\ri \infty} \hspace{-3mm}{\rm d}s
\int_{\gamma'-\ri \infty}^{\gamma'+\ri \infty}\hspace{-3mm} {\rm d}s'
\Gamma(s) \Gamma(s') 
\Gamma(1-s-s') \nn \\
& &\times \frac{y^{s'-1}}{(1-y)^{s+s'}}
\frac{(-m_1^2m_3^2+\e)^{-s}}{(m_2^2m_4^2+\e)^{-s}}
\frac{(st+\e)^{-s'}}{(m_2^2m_4^2+\e)^{-s'}}.
 \label{eq:f56}
\end{eqnarray}
If we now compare this expression with the expressions
(\ref{eq:f41}) and 
(\ref{eq:f45}), with $y_1=1/2$, we get
\begin{eqnarray}
\lefteqn{K(s, t, m_1^2, m_3^2, m_2^2, m_4^2)=-2\ri\pi\,{\rm sign}(\epsilon)
\frac{\ri }{(4\pi)^2}} \label{eq:f57}\\
& &{}\hspace{-3mm}\times \int_0^1\!\!{\rm d}y 
\big[\,y\,m_2^2m_4^2+(1-y)\,st -y (1-y)\,m_1^2m_3^2 +\e\,\big]^{-1}.\nn
\end{eqnarray}
Next, after employing the identity 
\begin{eqnarray}
\frac{1}{y\,m_2^2m_4^2+(1-y)\,st-y (1-y)\,m_1^2m_3^2+\e}& &\nn \\
=\frac{1}{\lambda}\left[
\frac{1}{y-(y_1-\e)}-\frac{1}{y-(y_2+\e)}\right],& &\label{eq:f58}
\end{eqnarray}
where 
\[ y_1=\frac{m_1^2 m_3^2-m_2^2 m_4^2+s t+\lambda}{2 m_1^2 m_3^2}\]
and
\[ y_2=\frac{m_1^2 m_3^2-m_2^2 m_4^2+s t-\lambda}{2 m_1^2 m_3^2}\]
are the roots of the equation 
\[ y\,m_2^2m_4^2+(1-y)\,st-y (1-y)\,m_1^2m_3^2=0,\]
while
\begin{eqnarray}
\lambda&=&\big[ (s t)^2+(m_1^2 m_3^2)^2+(m_2^2 m_4^2)^2
-2\,s t m_1^2 m_3^2 \nn \\
 & &{}
-2\,s t m_2^2 m_4^2-2\,m_1^2 m_3^2 m_2^2 m_4^2\big]^{1/2},\label{eq:f62}
\end{eqnarray}
and performing the remaining integration, 
we arrive at the final result for the
correction term:
\begin{eqnarray}
\lefteqn{K(s, t, m_1^2, m_3^2, m_2^2, m_4^2)=
-2\ri\pi\,{\rm sign}(\epsilon)\frac{
\ri }{(4\pi)^2\lambda}}\nn \\
& &{}\times \left[ 
\ln\left( m_2^2 m_4^2 + s t - (m_1^2 m_3^2-\lambda)(1+\e)\right)\right.\nn \\
& &{}\left.-
\ln\left( m_2^2 m_4^2 + s t - (m_1^2 m_3^2+\lambda)(1-\e)\right)
\right].\label{eq:f59}
\end{eqnarray}
As it is seen, the correction term $K$ is written in the form in which the
symmetry with respect to the pairs $(s,t)$ and $(m_2^2, m_4^2)$ is evident.

Based on the above consideration, 
the expression for the IR finite one-loop scalar box integral can be written as
\begin{eqnarray}
I_4(s,t;m_1^2,m_2^2,m_3^2,m_4^2,\epsilon)&=&
I_3(s t, m_1^2 m_3^2, m_2^2 m_4^2,\epsilon) \nn \\
& &{}\hspace{-2cm}+K(s,t,m_1^2,m_3^2,m_2^2,m_4^2,\epsilon).\label{eq:f61}
\end{eqnarray} 
The first term on the right-hand side in (\ref{eq:f61}) represents the
expression for the IR finite triangle scalar integral. It is evaluated 
in detail in the
Appendix, and is given by
\begin{eqnarray}
\lefteqn{\hspace{-5mm}I_3(\mu_1^2,\mu_2^2,\mu_3^2,\epsilon)=\frac{
\ri }{(4\pi)^2}\,
\frac{1}{\nu_3 (x_1-x_2)}\, 
\Bigg\{ \, 2\, \mbox{Li}_2\left( \,\frac{1}{x_2}\,\right)}\nn \\
& &{}-2\, \mbox{Li}_2\left( \,\frac{1}{x_1}\,\right)+
\ln \left[ \,x_1 x_2+\e \,{\rm sign}(\nu_3) \,\right] \nn \\
& &{}\times \left[ 
\,\ln \left( \frac{1-x_1}{-x_1}\right)
-\ln \left(\frac{1-x_2}{-x_2}\right) \,\right]\Bigg\},\label{eq:f35}
\end{eqnarray} 
where $x_1$ and $x_2$ are the roots of the equation
\[ x\,\nu_1+(1-x)\,\nu_2-x(1-x)\,\nu_3=0,\]
and
\[ \{ \nu_1,\nu_2,\nu_3\}={\cal P}\{ \mu_1^2,\mu_2^2,\mu_3^2\},\]
where ${\cal P}$ denotes the permutation of the masses 
chosen in a such a way that the roots 
$x_1$ and $x_2$ are outside
the interval $[0,1]$.
If the signs of all the masses 
$\mu_i^2$ are the same,
$\nu_3$ should be chosen so as to correspond to the mass 
with the smallest absolute value. 
On the other hand, if the signs of the masses are not all the same,
$\nu_3$ ought to
be chosen so as to coincide with the mass whose sign is opposite to 
the sign of the
other two masses.

The result (\ref{eq:f35}) is similar to the result
given in Ref. \cite{davy}, but is more convenient for numerical calculations
since it is easier to separate the real and imaginary part.
Note that as a consequence of choosing $\nu_i$ in an
appropriate way, the causal
$\e$ appears only in one of the logarithms in (\ref{eq:f35}).

The second term on the right-hand side in (\ref{eq:f61}) is the correction term.
As stated earlier,
it accounts for the fact that in the product of two kinematic variables we lose
information about the phases which can lead to missing some poles. 
The expression for the correction term given in 
(\ref{eq:f59}) corresponds to a special case characterized by the fact that the
$m_1^2, m_3^2 < 0$.
This term, however, can be written in a general and 
for practical purposes more
convenient form:
\begin{eqnarray}
\lefteqn{K(\alpha_1,\beta_1,\alpha_2,\beta_2,\alpha_3,\beta_3,\epsilon)=}\nn \\
& &{}\frac{\ri }{(4\pi)^2}\frac{-2\ri\pi\,{\rm sign}(\epsilon)}{\lambda}
\sum_{i=1}^{3} \theta(-\alpha_i)\,\theta(-\beta_i)\nn \\
& &\times 
\left[
\ln\left( \sum\nolimits_{j\neq i} \alpha_j \beta_j-(\alpha_i \beta_i-\lambda)(1+\e)\right)\right.\nn \\
& &{}-\left.
\ln\left( \sum\nolimits_{j\neq i} \alpha_j \beta_j-(\alpha_i \beta_i+\lambda)(1-\e)\right)\right],
\label{eq:f63}
\end{eqnarray}
with $\lambda$ given in (\ref{eq:f62}). 
It should be observed that in practice at most one of the
terms in the sum in (\ref{eq:f63}) is different from zero.

The above result for the IR finite scalar box integral, as stated earlier,
has been derived under the assumption that the number of
positive kinematic variables is larger than the number of negative ones.
If the opposite is true, then owing to the symmetry 
\begin{eqnarray}
\lefteqn{I_4(s,t;m_1^2,m_2^2,m_3^2,m_4^2,\epsilon)}\nn \\
& &=I_4(-s,-t;-m_1^2,-m_2^2,-m_3^2,-m_4^2,-\epsilon),
\end{eqnarray}
which is obvious from (\ref{eq:f6}), the
corresponding expression can simply be obtained from the one given 
above simply by changing the
signs of all six kinematic variables as well as of the causal $\epsilon$.

Finally, we have numerically checked our result for the IR finite scalar box
integral with massless internal lines with the corresponding result of Ref.
\cite{denner} and found agreement in all the kinematic regions.

\section{Conclusion}
\label{sec:con}
Using the Mellin-Barnes integral representations, the calculation of the
IR finite one-loop box scalar integral with massless internal lines
has been reduced to the calculation of the IR finite triangle scalar integral
with massless internal lines.
The result is presented in a very simple and compact form 
(only two dilogarithms) and, owing to the fact
that we have kept the causal $\e$ systematically throughout the calculation,
is quite general, i.e. valid for arbitrary values of the relevant
kinematic variables.
From the practical point of view, the main advantage of our result over the
results
previously obtained is that the arguments of the occurring logarithms and
dilogarithms are very simple, which enables one to separate the
real and imaginary
part of the integral very easily.
This makes our result for the IR finite one-loop scalar box integral more
appropriate for numerical calculations.

\begin{acknowledgements}
This work was supported by the Ministry of Science and Technology
of the Republic of Croatia under Contract No. 00980102.
\end{acknowledgements}

\section*{Appendix}
In the Appendix we present the details of the calculation of the IR massless
scalar one-loop triangle integral given in (\ref{eq:f35}).

Eliminating the
$\delta-$function in (\ref{eq:A2}) by performing the $x_3$ integration, 
making the change of variables $x_2 \to (1-x_1) x_2$ and performing the 
$x_1$ integration, the integral (\ref{eq:A2}) 
becomes
\begin{eqnarray}
\lefteqn{I_3(\mu_1^2,\mu_2^2,\mu_3^2)=\frac{\ri }{(4\pi)^2}
\int_0^{1}\!{\rm d}x } \label{eq:A3}\\
& &\!\!\!\times \frac{\ln \left[ x\,\mu_1^2+(1-x)\,\mu_3^2+\e \right]
- \ln \left[ x (1-x)\,\mu_2^2+\e \right]}{ x\,\mu_1^2+(1-x)\,\mu_3^2 
- x (1-x)\,\mu_2^2},\nn
\end{eqnarray}
where we have replaced the variable $x_2$ by $x$. The poles of the integrand are
\begin{equation}
x_{1, 2}=\frac{1}{2}\left[ 1-\frac{\mu_1^2}{\mu_2^2}+\frac{\mu_3^2}{\mu_2^2}\pm
\sqrt{\left( 1-\frac{\mu_1^2}{\mu_2^2}-\frac{\mu_3^2}{\mu_2^2}\right)^2- 
4\frac{\mu_1^2}{\mu_2^2}\frac{\mu_3^2}{\mu_2^2}}\,\right].
\label{eq:A6}
\end{equation}
As can easily be seen from the denominator in (\ref{eq:A3}), the integration
path will not pass through the pole if one of the following two cases 
is realized: 
first, if the
sign of $\mu_2^2$ is opposite to the signs of  $\mu_1^2$ and $\mu_3^2$; second,
if all the masses are of the same sign and $\mu_2^2$ has the smallest absolute
value.
Now, owing to the fact that the integral
$I_3(\mu_1^2,\mu_2^2,\mu_3^2)$ is invariant under the permutations of the
parameters $\mu_i^2$, one can easily show that it is always possible to permute
the parameters $\mu_i^2$ in a way that the poles $x_1$ and $x_2$ are outside the
interval $[0,1]$. In trying to accomplish that, one of the
following four possibilities can arise: the poles are complex conjugate, 
both poles are in the interval $\langle-\infty,0\rangle$,
both poles are in the interval $\langle 1,\infty\rangle$ and one pole is
 in the interval $\langle -\infty,0\rangle$
and another in the interval $\langle 1,\infty\rangle$.
In the following considerations we assume that the parameters $\mu_i^2$ are
already chosen in 
such a way that the above requirement is satisfied.

Due to the fact that the poles belong to one of the above-mentioned cases,
it is convenient to rewrite the integral (\ref{eq:A3}) 
in terms of $x_1$ and $x_2$.
The integral (\ref{eq:A3}) then takes the form
\begin{eqnarray}
\lefteqn{I_3(\mu_1^2,\mu_2^2,\mu_3^2)=\frac{\ri}{(4\pi)^2}
\,\frac{1}{\mu_2^2}
\int_0^{1}\!{\rm d}x\frac{1}{(x-x_1)(x-x_2)}}\nn\\
& &\times \bigg( \ln \left[ 1+x\frac{1-x_1-x_2}{x_1 x_2}\right]
- \ln \left[ x (1-x)\right]\nn \\
& &{}+
\ln \left[ x_1 x_2+\e \,{\rm sign}(\mu_2^2)\right]\bigg).
\label{eq:f14}
\end{eqnarray}
Observe that the arguments of the first two logarithms are positive and
that there are no poles on the path of the integration.
Now, by the partial fraction decomposition the integral
(\ref{eq:f14}) can be written as 
\begin{eqnarray}
\lefteqn{I_3(\mu_1^2,\mu_2^2,\mu_3^2) = \frac{\ri}{(4\pi)^2}
\frac{1}{\mu_2^2(x_1-x_2)}}\nn \\
& &{}\times \left[ G_1(x_1, x_2)+
 G_2(x_1, x_2)+G_3(x_1, x_2) \right] ,
\label{eq:f15}
\end{eqnarray}
where
\begin{eqnarray}
G_1(x_1,x_2)&=&\int_0^1\!{\rm d}x 
\frac{1}{x-x_1} \ln \left[ 1+x \frac{1-x_1-x_2}{x_1 x_2}\right]\nn \\
& &{}\hspace{-4mm}-\int_0^1\!{\rm d}x 
\frac{1}{x-x_2} \ln \left[ 1+x \frac{1-x_1-x_2}{x_1 x_2}\right], \label{m1}\\
G_2(x_1,x_2)&=&-\int_0^1\!{\rm d}x 
\frac{1}{x-x_1} \ln \left[ x(1-x)\right]\nn \\
& &{}+\int_0^1\!{\rm d}x 
\frac{1}{x-x_2} \ln \left[ x(1-x)\right], \label{m2}\\
G_3(x_1,x_2)&=&\ln \left[x_1 x_2+\e \,{\rm sign}(\mu_2^2)\right]\nn \\
& &{}\times \left(\int_0^1\!{\rm d}x 
\frac{1}{x-x_1} -
\int_0^1\!{\rm d}x 
\frac{1}{x-x_2} \right). \label{m3}
\end{eqnarray}
Let us now evaluate the above integrals in turn.
Changing the integration variable
\[ 1+x \frac{1-x_1-x_2}{x_1 x_2} \to \frac{1-x_1}{x_2}x \] 
in the first, and
\[ 1+x \frac{1-x_1-x_2}{x_1 x_2} \to \frac{1-x_2}{x_1}x \] 
in the second integral in (\ref{m1}), one finds
\begin{eqnarray}
G_1(x_1,x_2)&=&\int_{\frac{x_2}{1-x_1}}^{\frac{1-x_2}{x_1}}\!{\rm d}x 
\frac{1}{x-1} \ln \left[\frac{1-x_1}{x_2}x\right]\nn \\
& &{}\hspace{-4mm}-\int_{\frac{x_1}{1-x_2}}^{\frac{1-x_1}{x_2}}\!{\rm d}x 
\frac{1}{x-1} \ln \left[\frac{1-x_2}{x_1}x\right].
\end{eqnarray}
Next, after passing to the new integration variable
$x\to 1/x$ in the second integral above, making 
a small rearrangement and performing
two elementary integrations, the final expression for
the integral (\ref{m1}) turns out to be
\begin{equation}
G_1(x_1,x_2)=\frac{1}{2} \ln^2\left(\frac{1-x_1}{-x_1}\right)-
\frac{1}{2} \ln^2\left(\frac{1-x_2}{-x_2}\right).\label{m4}
\end{equation}
As for the integral (\ref{m2}), one can write it in the form
\begin{eqnarray}
G_2(x_1,x_2)&=&-\int_{0}^{1}\!{\rm d}x \frac{\ln (x)}{x-x_1}+
\int_{0}^{1}\!{\rm d}x \frac{\ln (x)}{x-(1-x_1)}\nn \\
& &\hspace{-6mm}{}+
\int_{0}^{1}\!{\rm d}x \frac{\ln (x)}{x-x_2}-
\int_{0}^{1}\!{\rm d}x \frac{\ln (x)}{x-(1-x_2)},
\end{eqnarray}
which, when expressed in terms of the Euler dilogarithms, takes the form
\begin{eqnarray}
G_2(x_1,x_2)&=&-\mbox{Li}_2\left( \frac{1}{x_1}\right)+
\mbox{Li}_2\left( \frac{1}{1-x_1}\right)\nn \\
& &\hspace{-1.5mm}{}+\mbox{Li}_2\left( \frac{1}{x_2}\right)-
\mbox{Li}_2\left( \frac{1}{1-x_2}\right). \label{m5}
\end{eqnarray}
Next, applying the identity
\[\mbox{Li}_2\left( \frac{1}{1-z}\right)=-\mbox{Li}_2\left( \frac{1}{z}\right)
-\frac{1}{2}\ln ^2\left(\frac{1-z}{-z}\right)\]
in (\ref{m5}), we arrive at
\begin{eqnarray}
G_2(x_1, x_2) &=&
-2~ \mbox{Li}_2\left( \frac{1}{x_1}\right)+
2~ \mbox{Li}_2\left( \frac{1}{x_2}\right)\nn \\
& &\hspace{-6mm}{}-
\frac{1}{2}\ln ^2\left(\frac{1-x_1}{-x_1}\right)+
\frac{1}{2}\ln ^2\left(\frac{1-x_2}{-x_2}\right). \label{m6}
\end{eqnarray}
The remaining integral (\ref{m3}) is elementary, and given by
\begin{eqnarray}
G_3(x_1,x_2)&=&\ln \left( x_1 x_2 +\e \,{\rm sign}(\mu_2^2)\right)\nn \\
& &\hspace{-6mm}{}\times \left[ \ln\left(\frac{1-x_1}{-x_1}\right)-
\ln\left(\frac{1-x_2}{-x_2}\right)\right].\label{m7}
\end{eqnarray}
Upon combining
Eqs.(\ref{eq:f15}), (\ref{m4}), (\ref{m6}) and (\ref{m7}),
we arrive at the final expression for the IR finite triangle scalar integral
given by Eq. (\ref{eq:f35}).


\end{document}